\documentstyle[12pt]{article}

\begin{document}
\topskip 2cm
\begin{titlepage}

\begin{center}


{\large\bf
CHARGED LEPTON G-2 AND CONSTRAINTS ON NEW PHYSICS
}\\
\vspace{2.5truecm}
{\large A.I. Studenikin}\footnote{\normalsize E-mail: studenik@srdlan.npi.msu.su}\\

\vspace{.5cm}

{\sl Department of Theoretical Physics, Physics Faculty,\\
 Moscow State
University, 119899 Moscow, Russian Federation}

\vspace{2.5cm}
\vfil

\begin{abstract}
A review of the theoretical and experimental values for the
charged lepton (electron and muon) anomalous magnetic moment
$a_l=\ (g_l-2)/2$
is presented. Employing the most accurate va\-lue for the fine structure
constant $\alpha^{-1}=\ 137.03599993(52) \ \ (0.0038\  ppm)$
obtained \cite {Kin196}
from the electron $(g-2)$ we find the new complete standard model
prediction for the anomalous magnetic moment of the muon
$a^{th}_{\mu}=\ 116591595(67)\times 10^{-11}$. The comparison of this
theoretical value and the precise experimental result \cite {Sch}
yields the estimation for the difference $\Delta a_{\mu}=\ a^{exp}_{\mu}- \
a^{th}_{\mu}$ at the $95\ \%$ confidence level:
$-95 \times 10^{-10}
\leq \Delta a_{\mu}\leq 236 \times 10^{-10}$.
The implication of the
expected a factor of about 20 increase of accuracy in the forthcoming
Brookhaven National Laboratory measurement of $a_{\mu}$ implies
$-47 \times 10^{-11}
\leq \Delta a_{\mu}\leq 118 \times 10^{-11}$, ($95\ \% \ C.L.$).
This interval is used to get constraints on the "new physics".
The value of the one-loop contributions $a^{B_i}_l$ of
different bosons predicted within extension of the standard model
and coupled to a charged lepton are discussed. The dependence of
$a^{B_i}_l$ on the masses of the bosons and leptons of the vacuum
polarization loops are investigated. The constraints on
"new physics" by requiring
that the new contributions $a^{B_i}_{\mu}$ to the muon anomalous
magnetic moment lie within the latter interval $\Delta a_{\mu}$ are
obtained.
\end{abstract}
\end{center}
\end{titlepage}

\eject


\section {Introduction}

There are two complementary approaches in exploration of
frontiers of particle physics.
One is based on experiments carried out at high energies on
accelerators and storage rings. At present, various programmes of
gaining on high energies are successfully realizing and even
more powerful accelerators are being constructed. However, it is
obvious that the traditional accelerators will not make it
possible in the future to sustain the present rate of advance
in experimental research towards
higher energies. That is why it is important to develop
another approach based on research of those
elementary
particle characteristics that can be measured in relatively low-energy
experiments and calculated theoretically with high accuracy. A comparison of
results of such experimental and high-precision theoretical studies
establishes a
non-accelerative method of getting information on properties of
elementary particles and their interactions.

	A unique example of such particle characteristics is provided
by the anomalous magnetic moments of charged leptons,
of the electron and the muon in particular. The
anomalous magnetic moment of a charged lepton is proportional to the deviation of the
so-called $g_{l}$-factor of the particle, that is the measure
of the size of the magnetic dipole
moment compared to its intrinsic angular momentum, from the value $g_{l}=g_{0}=2$.

Let me recall here the analogy with the classical motion of a charged
particle. Classically the magnetic dipole moment $\mu_{L}$ can arise when
a
charged particle is orbiting on the circular
trajectory with radius $r$. In this case the magnetic moment is
associated with the kinetic orbital momentum ${\vec L}={\vec r}\times {\vec p}$
and given by
${\vec \mu}_{L}=g_{cl}\ {e\over 2mc} {\vec L}$. Consequently, the $g$-factor for
this classical motion is equal to $g=g_{cl}=1$. Alternatively, for the
point-like Dirac particle the relation between the magnetic dipole moment
and the intrinsic spin moment takes the form

$${\vec \mu}_{m}=g_{l}\ {e_{l}\over{2m_{l}c}}{\vec S}, \eqno(1)$$
where ${\vec \mu}_{m}$ is the magnetic moment operator,
${\vec S}={\hbar\over2}{\vec \sigma}$ is
the spin operator, ${\vec \sigma}$ are the Pauli matrices, $e_{l}$ and $m_{l}$
are the charge and mass of the particle. As it follows, the $g$-factor
for this case is equal to

$$g_{l}=g_{0}=2. \eqno(2)$$

This value for the $g$-factor corresponds to the particular case when
the wave function of the particle obeys the Dirac equation and
interaction with external electromagnetic field is introduced via the
minimal coupling by extension of the derivative:
$\partial_{\mu}\rightarrow \partial_{\mu}+ieA_{\mu} $. This not trivial
result of the Dirac theory can be also received in the frame of the
non-relativistic approach based on the Pauli wave equation.

Note that the value $g_{l}=2$ represents a fundamental propriety of the
particle in respect to the electromagnetic interaction. If the particle
participate in any other interaction which endow it with an internal
structure then this departure from the point-likeness will be reflected
on the value of $g_{l}$-factor. For example, for the proton
$g_{p}$-factor is equal to 5.59 because of its internal structure.

It must be also mentioned that if the interaction with electromagnetic
field $F_{\mu\nu}$ is introduced in non-minimal way and the charged lepton wave
function obeys the equation
$$\big[(i\partial_\mu-eA_\mu)\gamma^\mu-m+{\Delta g\over 2}{e\over 4m}
\sigma^{\mu\nu}F_{\mu\nu}\big]\ \Psi=0,\ \
\sigma^{\mu\nu}={i\over 2}(\gamma^\mu \gamma^{\nu}-\gamma^{\nu}\gamma^{\mu})
,\eqno(3)$$
then the $g$-factor will be modified and equal to $g=2+\Delta g$.

However, even in the case of point-like Dirac particle as it was shown
first in \cite {Schw} the quantum nature
of minimal electromagnetic interaction shifts the value of
the $g_{l}$-factor. That is why it is convenient to represent the
charged lepton
$g_{l}$-factor in the form

$$g_{l}=2 \ (1+a_{l}), \eqno(4)$$
where $a_{l}$ is equal to the so-called anomalous magnetic moment measured
in units of the Bohr magneton $\mu^{0}_{l}$

$$a_{l}={\Delta \mu_{l} \over \mu^{0}_{l}}, \ \ \
\mu^{0}_{l}={e_{l}\hbar \over 2m_{l}c}. \eqno(5)$$

The most accurate experimental values for the electron
AMM was obtained by the Washington University group \cite {Sch}
and the experimental value for the muon AMM was obtained in CERN \cite {Bai}.
If one compares the observed value of the electron AMM with the
theoretical predictions one shall conclude that the main contribution
is due to quantum electrodynamical processes, while in the case of the
muon AMM the contributions of the strong and weak interactions are also
substantial. The latter still remain outside the limits of experimental
accuracy.
However, with expected more than factor of 20 improvement in forthcoming
results of the E821 Brookhaven National Laboratory measurement \cite {Hug}
of the muon AMM it
will become possible to detect the contributions of the weak
interactions.
Within the achieved accuracy the theoretical and experimental values
of the electron and
muon AMM are in good agreement with the predictions of the standard Glashow-
Weinberg-Salam model. This provides both a sensitive verification to
several orders in perturbation expansion of the QED and standard
$SU(3)_{C}\times SU(2)_{L}\times U(1)$ model as well as
put constraints on new
physics beyond the standard model \cite {KinMar,Kin196}. The expected improvements in accuracy
of measurements of the leptons AMM \cite {Hug,Dyck,Gab}
would either allow for the new physics effects to be
visible, or if no any extra contributions to the leptons AMM are seen,
more severe bounds on alternative models will be received.

The paper is organized as follows.
In the next Section 2 a brief review of the
experimental values of the electron and muon AMM are given.
In Sections 3 and 4 the present status of the theoretical values of the
leptons AMM is discussed and the new theoretical value for
the muon AMM is derived. In Section 5 we consider the present
discrepancy between the experimental
and theoretical values for the muon AMM and the one that could be achieved
in the near future.
The one-loop contributions of various types of bosons to
the charged lepton AMM are considered in Section 6 on the base of
calculation
of the subsequent contributions to the lepton mass operator in
external electromagnetic field. In Section 6 we also examine
the dependence of these
contributions to the lepton AMM on
masses of involved particles. In Conclusion we  derive
constraints on couplings of the muon to hypothetical bosons and on masses
of bosons.

\vskip5truemm
\section {Experimental values for anomalous magnetic $\ \ \ \ $
moments of electron
and muon}

The permanently increasing accuracy of theoretical
evaluations of charged leptons AMM is stimulated by the tremendous
accuracy that is achieved in measurements of the electron and muon
AMM. The latest experimental results \cite {Sch} of the Washington
University group for the electron and positron
AMM are \cite {Sch} $$a^{exp}_{e^{-}}=\ 1159652188.4(4.3)\times
10^{-12},\ \ a^{exp}_{e^{+}}=\ 1159652187.9(4.3) \times
10^{-12}.\eqno(6)$$ In these measurements a single electron and
positron are confined within a Penning trap which constrains the
position of a particle in a uniform magnetic field that is imposed by
means of a hyperboloid cavity. The experimental uncertainties are
dominated by a cavity shift effects of $\pm4\times 10^{-12}$, which
arises from a lack of control over the resonant interaction of a
particle with the electromagnetic modes of the surrounding microwave
cavity of the Penning trap \cite {Kin196}. There are also attempts to
reduce this uncertainty \cite {Dyck,Gab}.

The latest CERN storage ring measurements of the muon $\mu^{+}$ and $\mu^{-}$
AMM give the results \cite {Bail}
$$a^{exp}_{\mu^{-}}=\ 116593700(1200)\times 10^{-11},\ \
a^{exp}_{\mu^{+}}=\ 116591100(1100)\times 10^{-11},\eqno(7)$$
which together provide the current experimental average for the muon AMM
\cite {PDG},
$$a^{exp}_{\mu}=\ 116592300(840)\times 10^{-11}. \eqno(8)$$
The dominant error here is the statistical counting error of
$7 ppm$ in the
determination of the $(g_{\mu}-2)$ precession frequency which is the
difference between the spin precession frequency and the cyclotron
frequency of the muon in the magnetic field of the ring. It should
be possible to reduce this error \cite {KinH}
by at least a factor of 30 in the E821
BNL experiment because the intensity of the primary proton beam will
be a factor of about 100 greater and the storage ring magnetic field a
factor about of 3 greater than those used in the CERN experiment. The
reduction of the systematic error somewhat below the expected level
of the statistical error will be provided by the improvement of the
homogeneity of the magnetic field and diminishing of deviations of
muons orbits from the ideal reference orbit. Thus the present precision
the measurement of $a_{\mu}$ at the level of $84\times 10^{-10}$
is expected to be improved to the level of $4\times 10^{-10}$ or
even less ($\sim 1 - 2 \times 10^{-10}$) \cite {CKM,Bun}.

\vskip5truemm

\section {Anomalous magnetic moments of electron in standard model}

In the quantum electrodynamics the AMM of an electron (as well as of
the other charged leptons $l$) can be expressed as a perturbation series
expansion in the fine structure constant $\alpha$:

$$a_{l}=\sum\limits_{n}\Big({\alpha \over \pi}\Big)^{n}A^{(l)}_{n}+
\sum\limits_{n}\Big({\alpha \over \pi}\Big)^{n}B^{(l)}_{n}.\eqno(9)$$
The coefficients $A^{(l)}_{n}$ are independent of the lepton mass, thus
the first summ in (9) is identical for all charged leptons. The
coefficients $B^{(l)}_{n}=B^{(l)}_{n}\big({m_l \over m_{l'}}\big)$
are functions of ratios of the mass of the
external lepton $l$ to those of leptons $l'$
in the vacuum polarization loops.

In the case of the electron the mass independent terms has been evaluated
to order $\alpha ^{4}$:
$$a^{QED}_{e}(mass\ ind.)=A_{1}\Big({\alpha \over \pi}\Big)
+A_{2}\Big({\alpha \over \pi}\Big)^2+
A_{3}\Big({\alpha \over \pi}\Big)^3+
A_{4}\Big({\alpha \over \pi}\Big)^4, \eqno(10)$$
where
$$A_1=0.5,\  A_2=-0.328478965...,$$ $$ A_3=1.181241456..., \
 A_4=-1.4092(384).  \eqno(11)$$ The coefficients of the leading term
$A_1$ corresponds to the Schwinger value of the lepton AMM \cite
{Schw}. The coefficients $A_2$ and $A_3$ has been evaluated both in
numerical \cite {Kin196,Kin296,Kin95} and analytical \cite
{Pet}-\cite {LapRem} technique, whereas $A_4$ has been evaluated only
in numerical calculations \cite {Kin296}.

Three most important mass dependent terms in (9) for the electron AMM are known
analytically \cite {Elend,Lap,Kin196}
$$B_2\big(m_e/m_\mu\big)\Big({\alpha \over\pi}\Big)^2=
\ 2.804\times 10^{-12},$$
$$B_2\big({m_e/m_\tau}\big)\Big({\alpha \over\pi}\Big)^2=
\ 0.010\times 10^{-12}, \eqno(12)$$
$$B_3\big({m_e/m_\mu}\big)\Big({\alpha \over\pi}\Big)^3=
\ -0.924\times 10^{-13}.$$
Other QED contributions to the electron AMM are too small to be accounted
at present.

To get the theoretical value for the electron AMM, one must add also
contributions of the lowest-order ${\cal {O}}(\alpha^2)$
and of the order ${\cal {O}}
(\alpha^3)$ hadronic vacuum polarization and electroweak interaction
\cite {DH98,Kin196}:  $$a^{had}_e=1.635\times 10^{-12}, \
a^{EW}_e=0.030\times 10^{-12}. \eqno(13)$$

It has been shown that the
two-loop electroweak contributions to a charged lepton AMM are rather
important. The recent evaluation of the two-loop elecrtoweak
contributions to the electron AMM \cite {Kukh,PPR,CKM} amounts to
$-35\%$ of the one-loop term \cite {EW}.

Finally, from eqs. (10)-(13) it is possible to obtain the
theoretical value for the electron AMM \cite {Kin196} $$a^{th}_e=\
1159652156.4(1.2)(22.9)\times 10^{-12}, \eqno(14)$$ where the value
of $\alpha$ based on the latest measurements of the quantum Hall
effect \cite {Jeff} is used:
$$\alpha^{-1}(Hall)=\ 137.0360037(27)\ \ \ (0.020 \ ppm). \eqno(15)$$
The theoretical (14) and experimental
(6) values for the electron AMM are in agreement within the $1.4$
standard deviations level.

Note that the intrinsic theoretical uncertainty $1.2\times 10^{-12}$
of $a^{th}_e$ is much less than the uncertainty $22.9\times 10^{-12}$
from the measurements of $\alpha (Hall)$. It follows, that comparison
of the theoretical and experimental values of $a_e$ gives a more precise
value of the fine structure constant $\alpha$ than one used for the
derivation of $a^{th}_e$. The value of $\alpha$ determined from these
arguments with the use of the average of $a_{e^{-}}$ and $a_{e^{+}}$
is \cite {Kin196}
$$\alpha^{-1}_{(g_{e}-2)}=\ 137.03599993(52) \ \ \ (0.0038 ppm),
\eqno(16)$$
where the most of the error comes from the experimental uncertainty in
the measurements of $a_{e}$. We shall use this value for $\alpha$
when we get the value of the mun AMM.

\vskip5truemm
\section{Anomalous magnetic moment of muon in the standard model}

The theoretical prediction for the muon AMM can be also divided into
$$a^{th}_{\mu}=a^{QED}_{\mu}+a^{had}_{\mu}+a^{EW}_{\mu}.\eqno(17)$$
At present the QED contribution is known to order $\alpha^5$ \cite {muQED}
$$a^{QED}_{\mu}={\alpha \over \pi}C_1+\Big({\alpha \over \pi}\Big)^2C_2+
\Big({\alpha \over \pi}\Big)^3C_3+\Big({\alpha \over \pi}\Big)^4C_4+
\Big({\alpha \over \pi}\Big)^5C_5,\eqno(18)$$
where
$$C_1=\ 0.5,\ C_2=\ 0.765857381(51),\ C_3=\ 24.050531(140),$$
$$C_4=\ 126.02(42),\ C_5=\ 930(170),\eqno(19)$$
and in the calculation of the $\tau$ lepton loops was used
$m_{\tau}=1777\ MeV$.
Here coefficients $C_1$ and $C_2$ are known analytically whereas
the other are derived by numerical integration ( see ref.\cite {KinMar}
for a review of the calculations of $a^{QED}_{\mu}$).
Employing the value of the fine structure constant
$\alpha^{-1}_{(g_{e}-2)}=\ 137.03599993(52)$ determined from the
electron AMM \cite{Kin196} gives the new value$$a^{QED}_{\mu}=\ 116584705(2)\times
10^{-11}, \eqno(20)$$
for the QED contribution to the muon AMM. This value is $1\times 10^{-11}$
less than one used previously ( see, for example, \cite {KraZoc97,DH98}).

The muon AMM is more sensitive to processes at smaller distances than the
electron AMM because of the large mass scale $(m_{\mu} \gg m_e)$.
That is why the effects of strong and electroweak interactions are
much more important in the value of $a_{\mu}$ than in one of $a_e$.

The hadronic contributions to the lepton AMM arises from two effects:
hadronic vacuum-polarization and hadronic light-by-light scattering.
The hadronic vacuum-polarization contributions to $a_{\mu}$ can be evaluated
within the dispersion theory using the experimental data on the total
hadronic cross-section for the annihilation of electrons and positrons,
$\sigma_{e^{+}e^{-}\rightarrow hadrons}$,
and perturbative QCD for the very high
energies. The recent evaluation \cite {DH98} of the leading order
$\cal O$ $(\alpha / \pi)^2$ hadronic vacuum polarization
contribution,
$a^{had}_{\mu}(vac.pol.,\alpha ^2)=\ 6924(62)\times 10^{-11}$,
together with the non-leading order vacuum contribution \cite {ADH98,K97}
\ $a^{had}_{\mu}(vac.pol,\alpha ^3)=\ -100(6)\times 10^{-11}$
give
$$a^{had}_{\mu}(vac.pol.)=\ 6824(62)\times 10^{-11}. \eqno(21)$$
Following ref.\cite {DH98} we use for the light-by-light scattering
contribution the overage of results of refs. \cite {HKin98,Bij96}
that is
$$a^{had}_{\mu}(\gamma \times \gamma)=\ -85(25)\times 10^{-11}.
\eqno(22)$$ The combination of (21) and (22) yields the final result
for the hadronic contribution to the muon AMM $$a_{\mu}^{had}=\
6739(67)\times 10^{-11}.\eqno(23)$$
Note that the principal contribution as well as the error come from the
hadronic vacuum polarization effects from the low energy region and the
improvement in $e^{+} e^{-}$ data near the $\rho$ meson resonance energy
$\leq 1\ GeV$ could significantly reduce the uncertainty.

The one-loop electroweak $a^{EW}_{\mu}(1\ loop)$
contributions to the charged lepton AMM have been
calculated in \cite {EW} (see also \cite
{TRSt83,TRSt89,St190,OPK86} for the evaluation of the one-loop
$Z$, $W$, and Higgs bosons contributions to a charged lepton AMM exactly accounting
for mass parameters $({m_{l}/M_B})^2$, where $M_B= M_Z,\ M_W$ or
$M_{Higgs}$):  $$a^{EW}_{\mu}(1\ loop)={5 \over 3}{G_{\mu}m_{\mu}^2
\over 8\sqrt{2}\pi^2} \Big[1+{1 \over 5}(1-4\sin^2\theta_W)^2+$$
$${\cal {O}}\Big(
{m^2_{\mu} \over M_W^2}\Big)+
{\cal {O}}\Big({m^2_{\mu} \over M_Z^2}\ln {M_Z^2 \over m_{\mu}^2}\Big)+
{\cal {O}}\Big({m^2_{\mu} \over M_{Higgs}^2}\ln {M_{Higgs}^2 \over m_{\mu}^2}
\Big)\Big]\approx \ 195 \times 10^{-11},\eqno(24) $$
where $G_{\mu}=1.16639(1)\times 10^{-5}\ GeV^{-1}$,
and $\sin^2\theta_W=\ 0.224$. Note the presence of large logarithmic
terms $\ln (M^2_B/m^2_\mu)$ in the expansion of the neutral bosons
contributions $a^{Z,\ H}_{\mu}$ over parameters $m^2_\mu/M^2_B$.

The consideration \cite {Kukh,CKM,PPR}
of the two-loop electroweak contributions to the muon AMM
leads to an overall $22.6\% $ reduction of $a^{EW}_{\mu}$
for $M_{Higgs}\approx 250 \ GeV$.
Thus, the present prediction for the electroweak contribution to the muon
AMM is \cite {CKM}
$$a^{EW}_{\mu}=\ 151(4) \times 10^{-11}.\eqno(25)$$
The error is due to uncertainties in $M_{Higgs}$ and quark two-loop effects,
the current uncertainty in $\theta_W$ is about $0.05\times 10^{-11}$
\cite {KinMar}.

Summing the different contributions (20), (23), and (25) and adding
errors in quadratures, one can obtain for the theoretical value of
the muon AMM
$$a^{th}_{\mu}=\ 116591595(67)\times 10^{-11}.\eqno(26)$$
This our result for $a^{th}_{\mu}$ is less than one of
ref.\cite {ADH98} because the new value \cite {Kin196} for $\alpha_{(g-2)}$
is used for the evaluation of the QED contribution (20).

\vskip5truemm
\section {Discrepancy of experimental and theoretical $\ \ \ \
\ $ values of muon anomalous magnetic moment}

The uncertainty in (26) is mostly dominated by the hadronic lowest
order contribution. The uncertainties in the QED (20)
and electroweak (25) terms, $\Delta^{QED}_{\mu}$ and $\Delta^{EW}_{\mu}$,
contribute on the level of a few percent that of hadronic term,
$\Delta^{had}_{\mu}$, :
$$\Delta^{QED}_{\mu}+\Delta^{EW}_{\mu}\ <\ 10\% \
\Delta^{had}_{\mu}.\eqno(27)$$

Remarkably, within the recently achieved improvement in computation of the
hadronic contribution $a^{had}_{\mu}$ the overall error in $a^{th}_{\mu}$
becomes about 2.3 times less than the contribution of the electroweak
interaction $a^{EW}_{\mu}$. Together with the anticipated precision of the
E821 experiment the further reduction of errors in $a^{had}_{\mu}$ by
a factor of about 2 will allow a direct detection of the electroweak
contribution to $a_{\mu}$ and also will provide a test of new physics
at the multi-TeV scale.

In the difference of the experimental (8) and
theoretical (26) values
$$\Delta a_{\mu}=\ a^{exp}_{\mu}-a^{th}_{\mu}=\ 705(843)\times10^{-11}
\eqno(28)$$
the dominated uncertainty is due to the experimental error.
From (28) we can derive the 95 $\%$ confidence level limits on $\Delta a_{\mu}$:
$$-95\times 10^{-10} \leq \ \Delta a_{\mu}\leq
\ 236 \times 10^{-10}.\eqno(29)$$

Let us
suppose (see also ref.\cite {KraZoc97}) that after the expected a factor
of about 20 increase of accuracy in the E821 experiment and the
further improvement in the calculation of the hadronic contribution
the total uncertainty in $\Delta a_{\mu}$ will be again due to the
experimental error. Then in the case when the central value for the
deviation $\Delta a_{\mu}$ will be also shifted down by a factor of
20 (i.e., it is supposed that the deviations from the standard model
will not be visible on the new level of accuracy)
we can obtain the new limits $$-47\times 10^{-11}\leq \ \Delta
a_{\mu}\leq \ 118\times 10^{-11}\eqno(30)$$ at the 95 \% confidence
level. These limits are used below in evaluation of constraints on
interactions of the muon with hypothetical bosons that could be
received from the forthcoming data of the BNL E821 experiment.

\vskip5truemm

\section {Different bosons contributions to charged lepton anomalous
magnetic moment}

In the various generalizations of the standard
$SU(3)_{C}\times SU(2)_{L}\times U(1)$ model (such as, e.g.,
grand unified theories, technicolor models, composite models,
models with horizontal symmetry, superstrings, etc)
a rich spectrum of new bosons is predicted. The couplings of these
new bosons $B_{i}$ to a charged lepton $l$ give the "new physics"
 contributions
$a_{l}^{B_{i}}$ to the lepton AMM via vacuum polarization loops.
Constraints on
the "new physics" can be obtained by requiring that the new contributions
$a^{B_i}_{l}$ to the lepton AMM lie within the discrepancy
$\Delta a_{l}$ of the experimental and theoretical values.

\vskip5truemm
\subsection {Evaluation of different bosons contributions
to anomalous magnetic moment of charged lepton}

Let us briefly describe evaluation
of various types of bosons contributions to a charged
lepton AMM. The most systematic way to calculate contributions
of various types of bosons $B_i$ to the lepton AMM is based on
consideration \cite {St190,St290}
of radiative corrections to the motion of the lepton in external
electromagnetic field that appears due to the corresponding interactions
of the lepton with bosons. We consider an equation analogous to the Schwinger
equation in quantum electrodynamics, which describes the motion of the
lepton in the presence of the electromagnetic field taking radiative
$B_i$ boson effects into account
$$\big[(i\partial_\mu-eA^{ext}_\mu)\gamma^\mu-m\big]\ \Psi (x)
=\ \int M^{B_i}(x',x)\Psi (x')dx' ,\eqno(31)$$
where $A^{ext}$ is the 4-potential of the external electromagnetic field
and $M^{B_i}$ is
the contribution to the lepton mass operator due to the
corresponding vacuum polarization effects.
As a starting point, we chose the
external field to be constant crossed electric $\vec E$ and magnetic $\vec B$
field ($\vec E \perp \vec B,\ E=B$), which in the case of relativistic
particles provides a good model for any constant electromagnetic
field ( a detailed discussion of this statement can be found in refs.
\cite {Rit69,Rit72} .

We consider the lowest-order contributions to the mass operator of
a charged lepton, i.e. contributions of different virtual processes of
the form:$$l\rightarrow \ l'  +\ B_i \rightarrow \ l.\eqno(32)$$
The lepton
$l' $ in the virtual polarization loop may be not identical to the
initial and final charged lepton $l$ (that is the case of the QED and
the neutral couplings of the standard mode) which corresponds
not only to the charged
couplings $l^{\pm}-\nu_l-W^{\pm}$ of the standard model but, for example,
to the
flavour-changing neutral couplings $l^{\pm}- l'^\pm-B^{0}_{l}$
that are predicted
in different models with the horizontal symmetry \cite {St190} or
in the two-Higgs doublet extensions of the standard model
\cite {KraZoc97,NSher98}.
Note that the lowest-order slepton-photino and wino-sneutrino
contributions of the SUSY extension of the standard model ( see
\cite {ArNath84,St191} and references therein) are also of
this type.

Let us specify the properties of the bosons $B_{i}$ and their couplings
to leptons. We assume that interactions of a charged lepton $l$ with
various bosons
$B_i$ are given by the Lorentz-invariant couplings
$${\cal L}_{l}^{B_i}
=\ g_{i}\bar {\psi}_{l}\Gamma_{i}\psi_{l' }\phi_{B_i},\eqno(33)$$
where the index $i=\ 1,\ 2,...,\ 6$ enumerates different types of
interaction and corresponding different types of virtual processes
(32) (for definiteness we consider the negatively charged lepton
$l^{-}$):  $$\Gamma_1=\ 1,\ \Gamma_2= \gamma_5,\ \Gamma_3=\
\gamma_{\mu},\ \Gamma_4=\ \gamma_{\mu} \gamma_5, \ \Gamma_5=\
\gamma_{\mu}, \ \Gamma_6=\ \gamma_{\mu}\gamma_5 .\eqno(34)$$ We also
suppose that the other partner of $B_i$ in the vacuum polarization
loop of the process (32) for the cases of the scalar and
pseudoscalar neutral bosons, $B_1=\ S^0,$ $ B_2=\ P^0$, could be
arbitrary charged leptons $l'^{-}$ with mass not indispensably equal
to the mass of the initial lepton $l^{-}$ ($m_{l}\not=\ m_{l'}$).

In the cases
of the vector and axial vector neutral bosons, $B_3=\ V^0,$ $B_4=\
A^0$, it is supposed that leptons $l'^{-}$ in the vacuum polarization
loops are
equal to the initial leptons $l^{-}$, $m_l =\ m_{l'}$.
In the cases of the charged
bosons, $B_5=\ V^{-}$ and $B_6=\ A^{-}$, the virtual leptons are the
massless neutrinos $l' = \nu_{l}$, $m_{l'}=\ 0$. This choice of pairs
of particles in the vacuum polarization loops of processies (32)
allows to investigate different contributions to the charged lepton
AMM not only in the standard electroweak model but also in the
supersymmetrical extensions and in a wide class of other alternative
models.

To the one-loop order the contributions to the mass operator of a
charged lepton is given by
$$M^{(2)}_i(x',x)=\ -ig^2_i\Gamma_iG_{l_i}(x',x)\Gamma_iD_{B_i}(x',x),
\eqno(35)$$
where $G_{l_i}(x',x)$ and $D_{B_i}(x',x)$ are propagators of the lepton
$l'_i$ and boson $B_i$ accounting for the crossed electromagnetic field
if $l'_i$ or $B_i$ are charged particles. The expressions for the charged
lepton and boson propagators in the crossed external electromagnetic
field are given in refs.\cite {Rit72,St290} . It is possible
to find the contribution $M^{(2)}_i(x',x)$ to the mass operator of the
lepton in the presence of the electromagnetic field in the so-called
$E_p(x)$ -representation \cite {Rit72} (that is used instead the momentum $p$
representation in calculations for the  zero electromagnetic field case):
$$M^{(2)}_i(x',x)=\ \int{dp'dp \over (2\pi)^8}E(p',x')
M^{(2)}_i(p',p)\bar E(p,x),\
\ \bar E(p,x)=\ \gamma _0E^{+}(p,x),\eqno(36)$$
where
$$E(p,x)=\ \big(1+e{\hat n \hat A \over 2np}\big)e^{-i(px-\eta(p,\phi))},
\ \hat n= \ n_\mu \gamma^\mu, \ \hat A=\ A_\mu \gamma^\mu,$$

$$\eta (p,\phi)=\ \int \limits_{0}^{\phi}
\big[{e^2 A^2 (\rho) \over 2np}-{epA(\rho) \over np}\big]d\rho,\ \phi=\ nx.
\eqno(37)$$
Here $A_\mu(\phi)$ is the 4-potential of the crossed field, the reference
frame is fixed by the condition
$n_\mu=\ (1,0,0,1)$, then $n^2=\ nA=\ 0$. The further details of calculations
are discussed in \cite {St290}, and for the contributions to
the AMM of a charged lepton $l$ of mass $m_1$, moving in the electromagnetic
field, we have got \cite {St190,St290,St191}
$$a_l^{B_i}(\chi)=\ {g_i^2 \over (2\pi)^2}\int \limits_{0}^{\infty}
{du \over (1+u)^3}\Big({u \over \chi} \Big)^{2 \over 3}
\ \Omega_i\Upsilon (z_i),
\eqno(38)$$
where
$$\Upsilon(z_i)=\ \int \limits_{0}^{\infty}\sin \big(z_ix+{x^3 \over 3}
\big)dx.$$
The arguments $z_i$ of the upsilon function $\Upsilon (z_i)$
for different contributions $a_l^{B_i}$ can be obtained from the
universal expression
$$z=\big({u \over \chi}\big)^{2 \over 3}
 \Big[-{1 \over u}+{1+u \over u}{m^2_2 \over m^2_1}
+{1+u \over u^2}{m^2_3 \over m^2_1}\Big]\eqno(39)$$
with the appropriate choice of the values for masses $m_{1,2,3}$.
Here $m_1$ denotes the mass $m_l$ of the charged lepton $l$, $m_2$ and $m_3$
stand for the masses of charged and neutral particles in the vacuum
polarization loop (32), correspondently.

\centerline{\bf {Table 1.}}

\begin{tabular}{| c || c | c | c | c | c | c | c |}
\hline
  &   &   &   &   &   &   &   \\
i & $B_i$ & $\Gamma_i$ & $l' _i$ & ${m_2 \over m_1}$ & ${m_3 \over m_1}$
& $z_i$ & $\Omega_i$  \\
  &   &   &   &   &   &   &   \\
\hline \hline
  &   &   &   &   &   &   &   \\
1 & $S^0$ & 1 & $l'^{-}$ & ${m_2  \over m_1}$ & ${m_3 \over m_1}$ &
 $z_1=z$ &
${1 \over 2}+{1+u \over 2}{m_2 \over m_1}$   \\
  &   &   &   &   &   &   &   \\
\hline
  &   &   &   &   &   &   &   \\
2 & $P^0$ & $\gamma_5$ & $l'^{-}$ & ${m_2  \over m_1}$ & ${m_3 \over m_1}$ &
 $z_2=z$ &
${1 \over 2}-{1+u \over 2}{m_2  \over m_1}$ \\
  &   &   &   &   &   &   &   \\
\hline
  &   &   &   &   &   &   &   \\
3 & $V^0$ & $\gamma_{\mu}$ & $l^{-}$ & 1 & ${m_3 \over m_1}$ &
 $z_3=({u \over \chi})^{2 \over 3}
 \big(1+{1+u \over u^2}{m_3^2 \over m_1^2}\big)$ & 1 \\
  &   &   &   &   &   &   &   \\
\hline
  &   &   &   &   &   &   &   \\
4 & $A^0$ & $\gamma_{\mu}\gamma_5$ & $l^{-}$ & 1 & ${m_3 \over m_1}$ &
 $z_4=({u \over \chi})^{2 \over 3}
 \big(1+{1+u \over u^2}{m_3^2 \over m_1^2}\big)$ &
 $-3-{4 \over u}-2u {m_1^2 \over m_3^2} $ \\
  &   &   &   &   &   &   &   \\
\hline
  &   &   &   &   &   &   &   \\
5 & $V^{-}$ & $\gamma_{\mu}$ & $l'^{0}$ & ${m_2\over m_1}$ & 0 &
 $z_5=({u \over \chi})^{2 \over 3}
 \big(-{1 \over u}+{1+u \over u}{m_2^2 \over m_1^2}\big)$ &
 $2+{1 \over u}-{1 \over u}{m_1^2 \over m_2^2}$  \\
  &   &   &   &   &   &   &   \\
\hline
  &   &   &   &   &   &   &   \\
6 & $A^{-}$ & $\gamma_{\mu}\gamma_5$ & $l'^{0}$ & ${m_2 \over m_1}$ &
0 & $z_6=({u \over \chi})^{2 \over 3} \big(-{1 \over u}+{1+u \over
 u}{m_2^2 \over m_1^2}\big)$ & $2+{1 \over u}-{1 \over u}{m_1^2
 \over m_2^2}$ \\
  &   &   &   &   &   &   &   \\
\hline \end{tabular}
$$  $$

In Table 1 columns denote, respectively,
the types, charges and exact ratios of masses
${m_2/m_1}$ and ${m_3/m_1}$
(if they are not arbitrary) of bosons and leptons
and the structure of their couplings $\Gamma_i$, as well as the arguments
$z_i$ and functions $\Omega_i$ which determine the integrand in Eq.(38).
Taking in mind the diversity in properties of scalar, $S^0$, and
pseudoscalar, $P^0$, bosons, that are predicted within different
alternative theoretical models, for these two cases we do not fix the
mass parameters $m_2/m_1$ and $m_3/m_1$.

The obtained result (38)
shows the dependence of $a_l^{B_i}$ on the characteristic
dynamical parameter
$$\chi=\ [-(eF^{\mu\nu}p_{\nu})^2]^{1 \over2}m_l^{-3},\eqno(40)$$
where $F^{\mu\nu}$ is the tensor of external electromagnetic field and
$p_\nu$ is the momentum of the charged lepton $l$.

Note that the analogous
representation in terms of the function $\Upsilon (z)$ of
photon contribution in the lowest order of QED to the electron AMM was
received in ref.\cite {Rit69} and derivation of the vector $Z$, $W$,
and scalar Higgs $H$ bosons contributions in the standard electweak model
can be found
in ref.\cite {St190} (see also references therein). The dependence of
the photon contribution to the AMM of the electron on the strength of
an external electromagnetic field was pointed out in ref. \cite {Gupta49}
and the dinamical nature of the AMM was demonstrated in ref.\cite {TerDor69}.

As it was mentioned above
the received formula (38) for the AMM of the charged lepton moving in
the crossed electromagnetic field gives also the expression for the AMM
of the lepton moving with relativistic energy in
a constant magnetic field.
In this case the dinamical field parameter
$\chi$ takes the form
$$\chi={B p_{\perp} \over B_0 m_l},\eqno(41)$$
where $p_{\perp}=\sqrt {2eBn}$ is the projection of the momentum of the
lepton on the plane perpendicular to the vector $\vec B$, n is the
Landau levels number in the magnetic field and $B_0=\ {m^2_{l}/e}$
is the critical magnetic field that for the case of the electron
is equal to $B_0^e=4.41\times 10^{13}\  Gauss$.                               .

In the discussed above experiments on the charged
leptons AMM the particles were moving  in the presence of the magnetic field.
The expected magnetic field induced shifts of contributions $a^{B_i}_l$
for the reasonable strength of the field are rather small and are not
accessible for observation in experiments on
measurements of the electron and muon AMM of the types that have been
already performed at CERN and the Washington University.
However, with the further increase
of accuracy in measurements of the electron and muon magnetic moments
the electromagnetic field dependence could become important.
To demonstrate the possible scale of influence of a magnetic field on the
value of the charged lepton AMM
let us consider \cite {StTer90} the conditions
that can be realized, e.g., at the Stanford Linear Collider. Suppose
that the typical
elecron beam energies are approximately $50\  GeV$ and the effective magnetic
fields around the electron bundle can approach $10^4 \ Gauss$. Under such
conditions the dynamical field correction to the electron AMM in the
lowest-order of QED can
reach the value of $\Delta a_e= \ 0.6\times 10^{-8}$ that exceeds the
electroweak and hadronic contributions as well as the $\alpha^4$ term
of the vacuum (i.e., field independent) QED contribution.

On the basis of expression (38) we can obtain \cite {St190,St290}
the asymptotic limits
of the  various contributions $a^{B_i}_l,\ B_i=\ S^0,\ P^0,\ V^0,$ \rm
$A^0,\ V^{-},\ A^{-}$,
to the AMM of a charged lepton
in electromagnetic field for small values of $\chi$ (that corresponds, for
example, to
the case of relatively weak magnetic fields, $B \ll B_0$, and not-too-large
energies) and large values of $\chi$ ( here we use the notation:
$\lambda=\ m^2_{B_i}/m^2_l$, where $m_{B_i}$ denotes masses of different
bosons $B_{i}$):
$$a^{S^0}_l (\chi)=\ {g_{1}^2 \over 4\pi}\Bigg\{{\
{1 \over \pi}\big[\ k_1+{\chi^2 \over 3\lambda^3}
+{\chi^2 \over \lambda^4}\big(\ln \lambda -{257 \over 60}\big)\big]\ ,\
 \  \  \ \chi \ll \lambda \atop
 \  \  \ {7\Gamma ({1 \over 3}) \over 18 \sqrt{3}(3\chi)^{2 \over 3}}\ ,
  \  \  \  \  \  \  \  \
\chi \gg \lambda^{3 \over 2}\ \  \  \   \  \  \ } \eqno(42)$$

$$a^{P^0}_l (\chi)=\ {g_{2}^2 \over 4\pi}\Bigg\{{\
{1 \over \pi}\big(\ k_2-{2\chi^2 \over 3\lambda^3}\big) ,\
 \  \  \ \chi \ll \lambda \ \ \ \ \atop
 \  \  \ -{5\Gamma ({1 \over 3}) \over 18 \sqrt{3}(3\chi)^{2 \over 3}}\ ,
  \  \  \  \  \  \  \  \
\chi \gg \lambda^{3 \over 2}\ \  \  \   \  \  \ } \eqno(43)$$

$$a^{V^0}_l (\chi)=\ {g_{1}^2 \over 4\pi}\Bigg\{{\
{1 \over \pi}\big[\ k_3
+2{\chi^2 \over \lambda^4}\big(\ln \lambda -{257 \over 60}\big)\big]\ ,\
 \  \  \ \chi \ll \lambda \atop
 \  \  \ {\Gamma ({1 \over 3}) \over 9 \sqrt{3}(3\chi)^{2 \over 3}}\ ,
  \  \  \  \  \  \  \  \
\chi \gg \lambda^{3 \over 2}\ \  \  \   \  \  \ }\eqno(44)$$

$$a^{A^0}_l (\chi)=\ {g_{1}^2 \over 4\pi}\Bigg\{{\
{1 \over \pi}\big[\ k_4+{\chi^2 \over \lambda^4}\big(-6\ln \lambda
+{691 \over 30}\big)\big]\ ,\
 \  \  \ \chi \ll \lambda \atop
 \  \  \ -{11\Gamma ({1 \over 3}) \over 9 \sqrt{3}(3\chi)^{2 \over 3}}\ ,
  \  \  \  \  \  \  \  \
\chi \gg \lambda^{3 \over 2}\ \  \  \   \  \  \ }\eqno(45)$$

$$a^{V^{-},\ A^{-}}_l (\chi)=\ {g_{5,6}^2 \over 4\pi}\Bigg\{{\
{1 \over \pi}\big(\ k_{5,6}+{\chi^2 \over 10\lambda^4}\big)\ ,\
 \  \  \ \chi \ll \lambda \ \ \ \ \atop
 \  \  \ {11\Gamma ({1 \over 3}) \over 9 \sqrt{3}(3\chi)^{2 \over 3}}\ ,
  \  \  \  \  \  \  \  \
\chi \gg \lambda^{3 \over 2}\ \  \  \   \  \  \ }.\eqno(46)$$
In evaluation of the terms depending on $\chi$ it was assumed that the
bosons $B_i$ are heavy and the condition $\lambda \gg 1$ holds.
We also set here the mass $m_2=\ m_1$ for the $S^0$ and $P^0$ bosons
contributions. However, the general case of $m_2 \not= m_1$ for these
contributions is considered below.

From eqs.(42)-(46) it follows that for any type of considered
bosons $B=\ S^0$, $\ P^0$, $\ V^0$, $\ A^0$, $\ V^{-}$, $\ A^{-}$
the dependence of the contributions to the
lepton AMM on parameter $\chi$ is the same. For $\chi \ll \lambda_i$ small
corrections quadratic in $\chi$ to the vacuum (i.e., field independent)
contributions $a^{B_{i}}_l(0)$ arise. At large $\chi$ ($\chi \gg
\lambda^ {3/2}$) all of the contributions explicitly
demonstrate the dynamic nature of the lepton AMM being proportional
to $\chi^{-{2/3}}$.  The similar behavior of the lowest-order
QED contribution to the lepton AMM was discovered in ref.
\cite {TerDor69,Rit69}.

The formulas (42), (44),
(45), and (46) allow us also to investigate the external field and energy
dependence of the standard model $Z$, $W$, and $H$ bosons
contributions to the charged lepton AMM.
The dependence of different contributions $a^{B_i}_l$ on the external
electromagnetic field and the lepton energy could reveal itself and
have to be accounted, for example,
in the case of the motion of relativistic leptons in the vicinity of
astrophysical objects like neutron stars, where strong magnetic fields
of the order $0.1\times B_0^e=\ 4.41\times 10^{12}\  Gauss$.
Under such conditions the values of
contributions as it follows from eqs.(42)-(46) could be substantially
less than the vacuum one-loop terms $a^{B_i}_l(0)$.

\vskip5truemm
\subsection {Dependence of bosons contributions to lepton anomalous
magnetic moment on masses of particles}

Let us now turn to the vacuum contributions to a charged lepton AMM,
 for which we obtain \cite {St290}
$$a^{B_l}_l=\ {g^2_i \over 8\pi^2}k_{B_i}(\lambda_i),\eqno(47)$$
where the functions
$k_{B_i}(\lambda_i)=\ k_i(\lambda)$ depend on the only one
mass parameter $\lambda_i={m^2_{B_i}/m^2_l}$:

$$k_1=\ \big({1 \over 2}\lambda^3-{5 \over2}\lambda^2+2\lambda\big)
\epsilon^{-1}\ln K_1+
({1 \over2}\lambda^2 -{3 \over 2}\lambda)
\ln\lambda-\lambda+{3 \over 2},\eqno(48)$$

$$K_i (\lambda_i)=\ K_i (\lambda)=\vert {\lambda-\epsilon \over
\lambda+\epsilon} \vert,
\ \epsilon= \vert\lambda(\lambda-4)\vert^{1 \over 2}, \  (\lambda \not=\ 4),
\eqno(49)$$

$$k_2=\ \big({1 \over 2}\lambda^2-{3 \over 2}\lambda \big)
\epsilon^{-1}\ln K_2+ ({1 \over 2}\lambda^2 -{1 \over 2}\lambda)
\ln\lambda-\lambda-{1 \over 2},\eqno(50)$$

$$k_3=\ \big(\lambda^3-4\lambda^2+2\lambda \big)
\epsilon^{-1}\ln K_2+ (\lambda^2 -2\lambda)
\ln\lambda-2\lambda+1,\eqno(51)$$

$$k_4=\ \big(\lambda^3-6\lambda^2+8\lambda \big)
\epsilon^{-1}\ln K_2+ (\lambda^2 -4\lambda +2)
\ln\lambda-2\lambda+5-{2 \over \lambda},\eqno(52)$$

$$k_5=k_6=\ \big(2\lambda^2-5\lambda+3 \big)
\ln L-2\lambda+4+{1 \over 2\lambda},\
L=\ \vert {\lambda \over \lambda-1}\vert\ . \eqno(53)$$

In Fig.1 the dependence on ratio $m_{B_i}/m_l$ of the functions
$k_i$, which determine the
vacuum contributions of the various types of bosons to the charged
lepton AMM are plotted. The number of the curve corresponds to the
value of index $i$. For instance, the curve (1) shows the dependence
of the function $k_1$ on $\sqrt \lambda_1$.
 The scalar, neutral vector, charged vector and
axial vector bosons contributions to the negatively charged lepton
are always positive, whereas the
contributions of the pseudoscalar and neutral axial vector bosons are
negative. The absolute values of all of the contributions go to zero
with the increase of the masses of the bosons.



For the cases of light ($m_{B_i}/m_l \ll 1$)
and heavy ($m_{B_i}/m_l \gg 1$) bosons it
is possible to obtain from eqs.(47)-(53) the limiting values for the
contributions to the negatively charged lepton AMM \cite {St290}.
If the mass ratio squared, $\lambda_i$, becomes small ($\lambda \ll 1$)
the functions
$k_i$ reduce to (here we again suppose that $\beta_i= \ m^2_{2(i)}/m^2_l=1$
for the scalar and pseudoscalar bosons contributions)

$$ k_1={3 \over 2},\ k_2=-{1\over 2},\ k_3=1, \\
\ k_4=-{2 \over \lambda} \rightarrow - \infty,\
k_5=k_6={2 \over \lambda} \rightarrow  \infty.\eqno(54)$$
In the limit of large $\lambda_i$ ($\lambda \gg 1$) the functions $k_i$ go to
zero like
$$ k_1={1 \over \lambda}\ln \lambda-{7 \over 6\lambda},\
k_2=-{1 \over \lambda}\ln \lambda+{11 \over 6\lambda},$$
$$k_3={2 \over 3\lambda}-{2 \over \lambda^2}\ln \lambda, \
k_4=-{10 \over 3\lambda}+{2 \over \lambda^2}\ln \lambda,\eqno(55)$$
$$ k_5=k_6={10 \over 3\lambda}+{2 \over 3\lambda^2}.$$
The expressions for the functions $k_{1,2,4}$ and their limiting
values correct mistakes that exist in papers of other authors (the detailed
discussion on this item see in refs.\cite {St291}).
As it follows from eqs.(54) the scalar, pseudoscalar, and neutral
vector bosons contributions remain finite in the limit of massless
bosons. Note that in this limit the vector boson contribution gives
the lowest-order QED result for the lepton AMM: $a^{QED}_l=\ \alpha /
2\pi,$ where $\alpha=\  e^2 / 4\pi$. On the contrary, the absolute
values of the neutral axial vector and charged vector and axial
vector bosons contributions increase to infinity in this limit.

It is worth to note that in the limit of heavy bosons the scalar and
pseudoscalar bosons contributions contain in addition to the
term $1/\lambda_i$ (which is common
for all of the contributions $a^{B_i}_l(0)$) an extra factor
of $\ln \lambda_i$ that can enlarge to some extent these two
contributions to $a_l$.

\vskip5truemm
\section{Conclusion}

Substituting of the specific values for the coupling constants $g_i$
and masses of the bosons $B_i$ to the expressions (47)-(53) one 
can get
the contributions
to the AMM of the charged lepton of mass $m_l$ that could arise in various
theories.
Thus, with the combination of formulas (48), (49), (51)-(53) for the $S^0,$ $V^0,$ $ A^0,$ $V^{-},$
and $A^{-}$ bosons contributions we can find the vacuum contributions
of the $Z,\ W,$ and $H$ bosons of the standard model (see ref.\cite {St190}
and references therein).

Formulas (47)-(53) for the one-loop bosons contributions
to the AMM of a charged lepton that exactly account for the masses
of virtual particles can be used for getting constraints on parameters 
of theoretical models.

Let us discuss bounds from the muon AMM data on the existence of hypothetical
forces which
would couple to the muon. Here we reexamine constraints of ref.\cite
{St291} and predict new limits on $\alpha_{B_{i}\mu}=\ g^2_i/4\pi$
and masses $m_{B_i}$ of hypothetical bosons $S^0,\ P^0,\ V^0$, and
$A^0$ that could be imposed by the improved measurements of the muon
AMM at the BNL.  The contributions of different bosons are considered
separately. To get constrains we demand that each of the new effects
could remove the discrepancy:  $$a^{B_{1,2,3,4,}}_{\mu}=\ \Delta
a_{\mu}.\eqno(56)$$ For $\Delta a_{\mu}$ we use the value of eq.(30).

The results as the exclusion plotts
are presented in figures 2-5 that show the constraints on
$\alpha_{B_{i}\mu}$ and $m_{B_i}$ in the decimal logarithmic scale.
The regions above curves are excluded. The curves on fig.2 and 4 set
the limits in the cases of bosons $S^0$ and $V^0$, respectively, and
are governed by the equation (see limits (30) on $\Delta a_{\mu}$)
$$\alpha_{B_{1,3}}=\ 2\pi\ 1.18\times 10^{-9}/
k_{1,3}(\lambda^{(\mu)}_{1,3}).\eqno(57)$$
The curves on fig.3 and 5
corresponds to the cases of bosons $P^0$ and $A^0$ and are governed
by the equation $$\alpha_{B_{2,4}}=\ -2\pi\ 0.47\times 10^{-9}/
k_{2,4}(\lambda^{(\mu)}_{2,4}).\eqno(58)$$









Comparing the curves in figs.2, 3, and 4 for the scalar, pseudoscalar,
and axial vector neutral bosons contributions with the similar curves
on figures of ref.\cite {St291} we can see that the constraints on
$\alpha_{B_i}$ for the fixed values of $m_{B_i}$
obtained here with use of the new bounds (30) are about one order of
magnitude as strong as those of ref.\cite {St291}.

The combined use of the $S^0$ and $P^0$ bosons
contributions enables us to get \cite {St191} the slepton-photino
and wino-sneutrino contributions 
to the lepton AMM in supersymmetric models.
The appropriate
combination of the $S^0$ and $P^0$ bosons contributions gives also
contributions to the lepton AMM in various models with the horizontal
symmetry \cite {St190}.

However, for
these two cases we need to know the vacuum contributions of the scalar
and pseudoscalar bosons in general form for arbitrary values of
the mass parameter $\beta_i=\
m^2_{2(i)}/m^2_l \not= \ 1$. This can be done with the help of
formulas (38), (39) and data of the first two lines of Table 1. For the
functions $k_1(\beta_1,\ \lambda_1)=\ k_1(\beta,\ \lambda)$ and
$k_2(\beta_2,\ \lambda_2)=\ k_2(\beta,\ \lambda)$ (here $\lambda_i=\
m^2_{3(i)}/m^2_l$) we get:
$$k_{1,2}(\beta,\ \lambda)=\ {1 \over 2}\bigg[\big((\lambda-\beta)^2
-\beta\big)\ln {\lambda \over \beta}+\bigg({(\lambda-\beta)^3 \over \beta}
-{\lambda ^2 \over \beta} -\lambda+2\beta-1\bigg)\epsilon'^{-1}\ln K'
+2(\beta-\lambda)-1\bigg]$$ $$\pm {\beta^{-{1 \over 2}} \over 2}\bigg[
{\beta-\lambda-1 \over \beta}\ln{\lambda \over \beta}-
\bigg(\big({\lambda \over \beta}-1\big)^2-{2 \over \beta} + {1 \over \beta^2}
\bigg) \epsilon'^{-1}\ln K'+{2 \over \beta}\bigg],\eqno(59)$$
where
$$\epsilon'=\ \vert {\rho^2-4{\lambda \over \beta}}\vert^{1 \over 2},\
\rho=\ 1+{\lambda-1 \over \beta},\ K'=\vert {\rho-\epsilon' \over
\rho+\epsilon'}\vert.$$

This two parametric ($\beta$, $\lambda$) analytic expression
can be used to derive the value the scalar and pseudoscalar
bosons contributions to the charged lepton AMM in different
theoretical models in which such spinless bosons are predicted.

\section{Acknowledgments}

It is a pleasure to thank Yury Chernyakov, Andrey Egorov and Rudolf Faustov
for discussions and assistance.
I should like to thank Mario Greco for inviting me to this interesting
Recontres and
all the organizers of the conference for their kind hospitality.

\eject

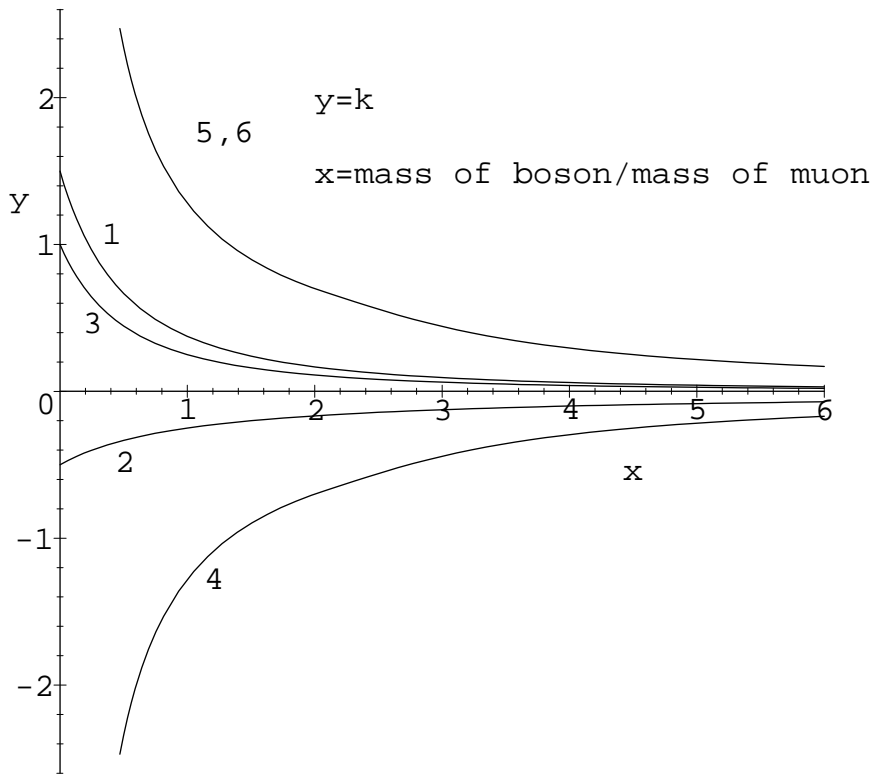
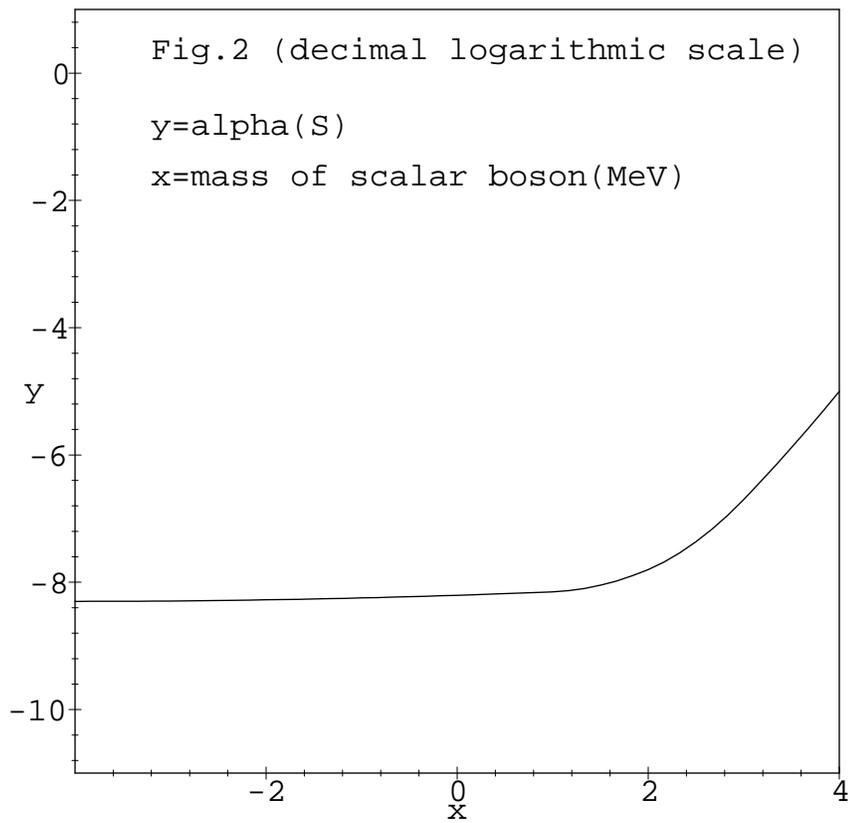
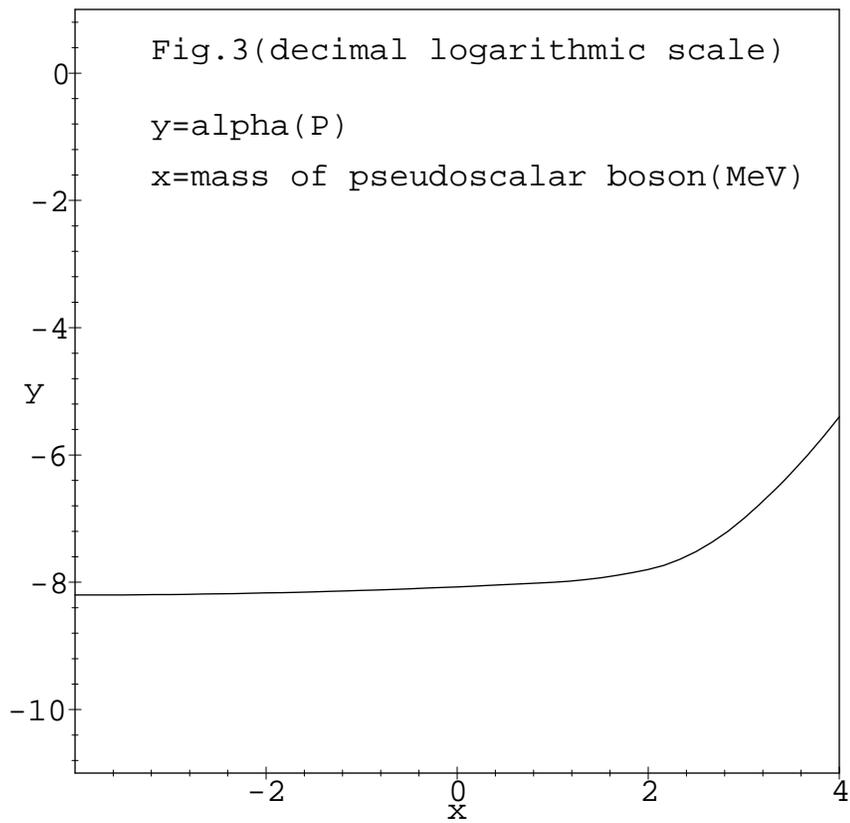
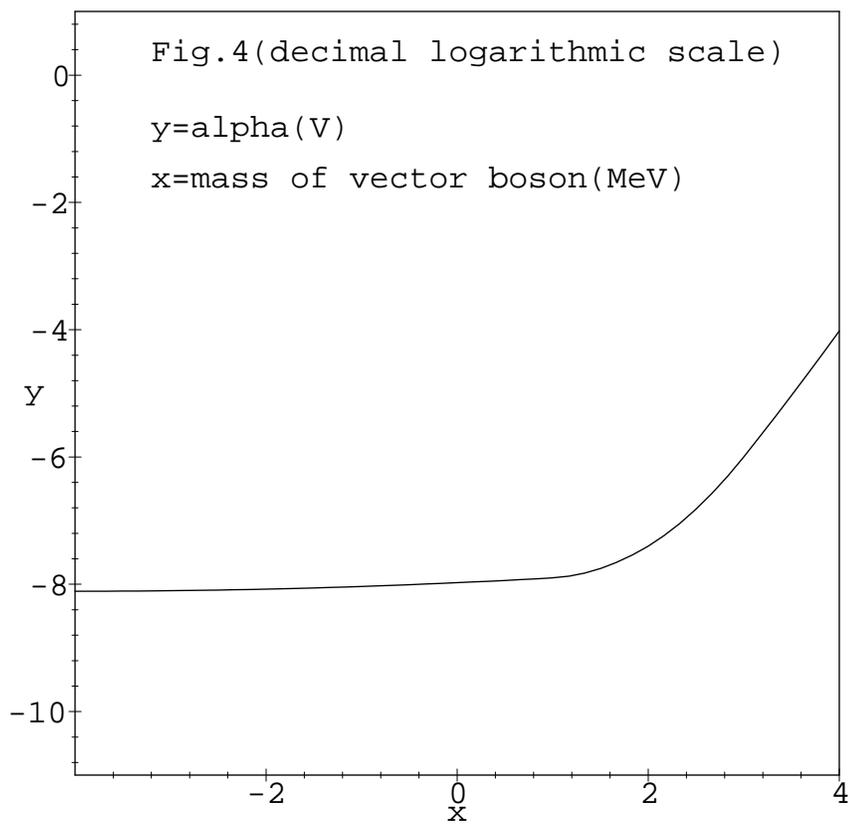
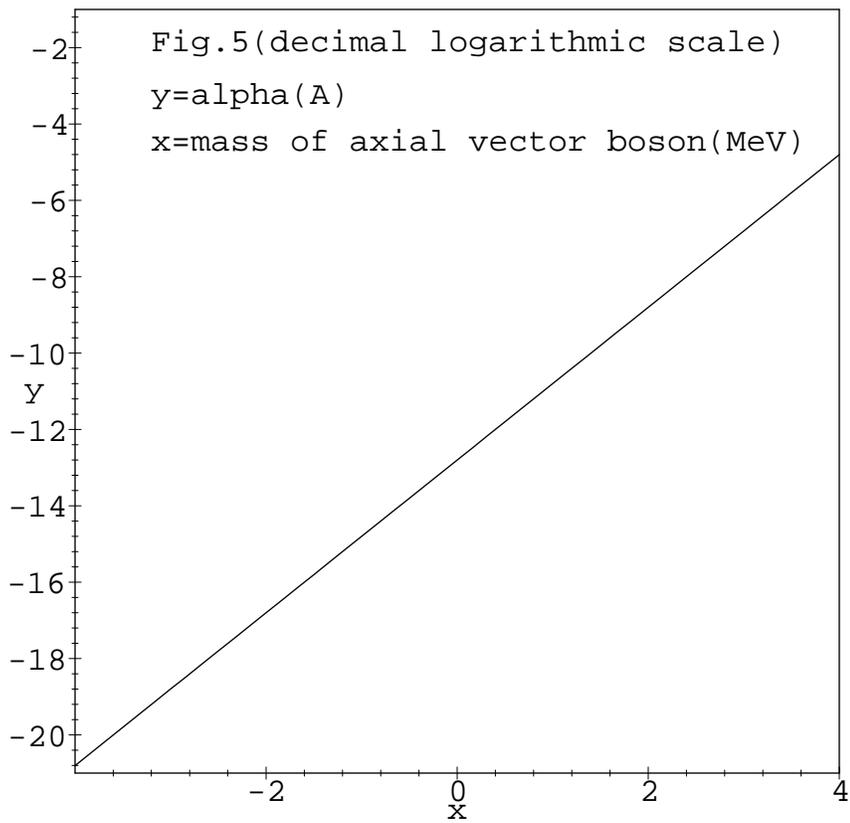
\end{document}